\documentclass[reprint,showpacs,preprintnumbers, aps, prd, longbibliography]{revtex4-1}

\eprint

\usepackage{graphicx, dcolumn, tabularx, bm, color, amsmath,amssymb}
\def\Ddots{\mathinner{\mkern1mu\raise\p@	
		\vbox{\kern7\p@\hbox{.}}\mkern2mu
		\raise4\p@\hbox{.}\mkern2mu\raise7\p@\hbox{.}\mkern1mu}}
\makeatother

\usepackage{lineno}
\begin{document}

	\title{	Adiabaticity parameters for the categorization of light-matter interaction -- from weak to strong driving}

	\author{Christian Heide} 
	\email{Christian.Heide@fau.de}
	\thanks{These two authors contributed equally}
	\affiliation{Laser Physics, Department of Physics, Friedrich-Alexander-Universit\"at Erlangen-N\"urnberg (FAU), Staudtstrasse 1, D-91058 Erlangen, Germany}
	\author{Tobias Boolakee}
	\email{Tobias.Boolakee@fau.de}
	\thanks{These two authors contributed equally}
	\affiliation{Laser Physics, Department of Physics, Friedrich-Alexander-Universit\"at Erlangen-N\"urnberg (FAU), Staudtstrasse 1, D-91058 Erlangen, Germany}
	\author{Takuya Higuchi}
	\affiliation{Laser Physics, Department of Physics, Friedrich-Alexander-Universit\"at Erlangen-N\"urnberg (FAU), Staudtstrasse 1, D-91058 Erlangen, Germany}
	\author{Peter Hommelhoff}
	\email{Peter.Hommelhoff@fau.de}
	\affiliation{Laser Physics, Department of Physics, Friedrich-Alexander-Universit\"at Erlangen-N\"urnberg (FAU), Staudtstrasse 1, D-91058 Erlangen, Germany}
	\date{\today}
	
	\begin{abstract}
		We investigate theoretically and numerically the light-matter interaction in a two-level system (TLS) as a model system for excitation in a solid-state band structure. We identify five clearly distinct excitation regimes, categorized with well known adiabaticity parameters: (1) the perturbative multiphoton absorption regime for small driving field strengths, and four light field-driven regimes, where intraband motion connects different TLS: (2) the impulsive Landau-Zener (LZ) regime, (3) the non-impulsive LZ regime, (4) the adiabatic regime and (5) the adiabatic-impulsive regime for large electric field strengths. This categorization is tremendously helpful to understand the highly complex excitation dynamics in any TLS, in particular when the driving field strength varies, and naturally connects Rabi physics with Landau-Zener physics. In addition, we find an insightful analytical expression for the photon orders connecting the perturbative multiphoton regime with the light field-driven regimes. Moreover, in the adiabatic-impulsive regime, adiabatic motion and impulsive LZ transitions are equally important, leading to an inversion symmetry breaking of the TLS when applying few cycle laser pulses. This categorization allows a deep understanding of driven TLS in a large variety of settings ranging from cold atoms and molecules to solids and qubits, and will help to find optimal driving parameters for a given purpose. 
	\end{abstract}%
	\maketitle
	
	
	The interaction between intense optical fields and two-level systems (TLS) has facilitated controlling electrons coherently on ultrashort timescales. This gave rise to new research areas including the efficient generation of high-harmonics in atoms and solids \cite{Corkum2013, Ghimire2019, Vampa2015, Jurgens2020}, light field-driven ionization and electron emission \cite{Ciappina2017} as well as light field-driven current generation in solids \cite{Schiffrin2013, Kelardeh2015, Higuchi2017}. 
	{Simultaneously, studying light-matter interaction in engineered electrodynamic environments led to the vibrant field of cavity quantum electrodynamics (QED) where recent systems based on superconducting qubits, exciton and intersubband polaritons and electron cyclotron resonance are used to explore their peculiar properties in the regimes of ultrastrong and deep strong coupling \cite{FriskKockum2019, FornDiaz2019}.} Likewise, coherent electron dynamics is investigated in qubits for information processing \cite{Berns2006, Otxoa2019} and, more recently, in topologically relevant materials \cite{Reimann2018, AzarOliaeiMotlagh2019, Vaswani2020, Nie2020}. 
	In all these cases, the underlying physics can be simplified, categorized and described with a TLS interacting with light. Here, we provide a clear categorization of different excitation regimes using well-known adiabaticity parameters, with a focus on solids. Finally we show that the presented categorization is applicable to a multitude of different systems.\\
	As long as the light-field is weak, momentum exchange between light and electrons, i.e., intraband motion, can be neglected, and the occupation of the bands is well described with resonant Rabi physics, with interband transitions only \cite{allen1987optical, cohen1998atom}. This picture, however, becomes unsuited or at least difficult to interpret when the light-field becomes so strong that different electron wave numbers become coupled resulting in field-driven intraband motion, which may influences interband transitions. In this case, the electron dynamics is better described as (repeated) Landau-Zener transitions between valence and conduction band \cite{Ashhab2007, Ishikawa2010, Shevchenko2010, Schiffrin2013, Kelardeh2015, Chizhova2016, Wismer2016, Fillion-Gourdeau2016, Higuchi2017, Kruchinin2018, Sato2018, Li2020}. An appropriate basis to describe intraband motion (coupling of different k-states) and interband transitions is the Houston basis \cite{Higuchi2017, Kruchinin2018}.\\	
	Whereas pure diabatic interband or adiabatic intraband electron dynamics have been well investigated in the context of strong-field physics, such as with high harmonic generation (HHG) in solids \cite{Corkum2013, Higuchi2014, Vampa2015, Luu2015, Liu2017, Tancogne-Dejean2017, Luu2018, Sato2018, Reimann2018, Kaneshima2018, Ghimire2019}, intraband motion and interband transitions acting in a combined fashion have been investigated to a less comprehensive degree. Based on adiabaticity parameters, we identify a novel regime of light-matter interaction, where both intraband motion and interband transitions generate an off-resonant, residual excitation. In contrast to HHG, which is not able to probe this off-resonant excitation directly, it can be well observed as a residual current \cite{Schiffrin2013, Kelardeh2015, Higuchi2017}. With this new regime, we can now give a complete and a general picture of TLS physics and can categorize it into five clearly distinct regimes.\\
	
	\begin{figure}[h!]
		\begin{center}
			\includegraphics[width=6cm]{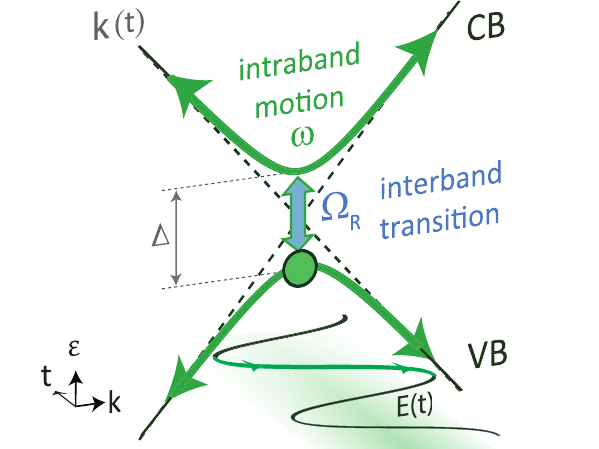}
			\caption{\label{Figure1} \textbf{Light-field driven electron dynamics in a two-level system.} When the electric field is small, intraband motion can be neglected and the excitation oscillates resonantly between valence (VB) and conduction band (CB) with the Rabi frequency $\Omega_\text{R}$. When the electric field becomes strong, the Rabi frequency exceeds the driving frequency, and intraband motion strongly influences interband transitions.} 
		\end{center}
	\end{figure}
	\begin{figure*}[t!]
		\begin{center}
			\includegraphics[width=17cm]{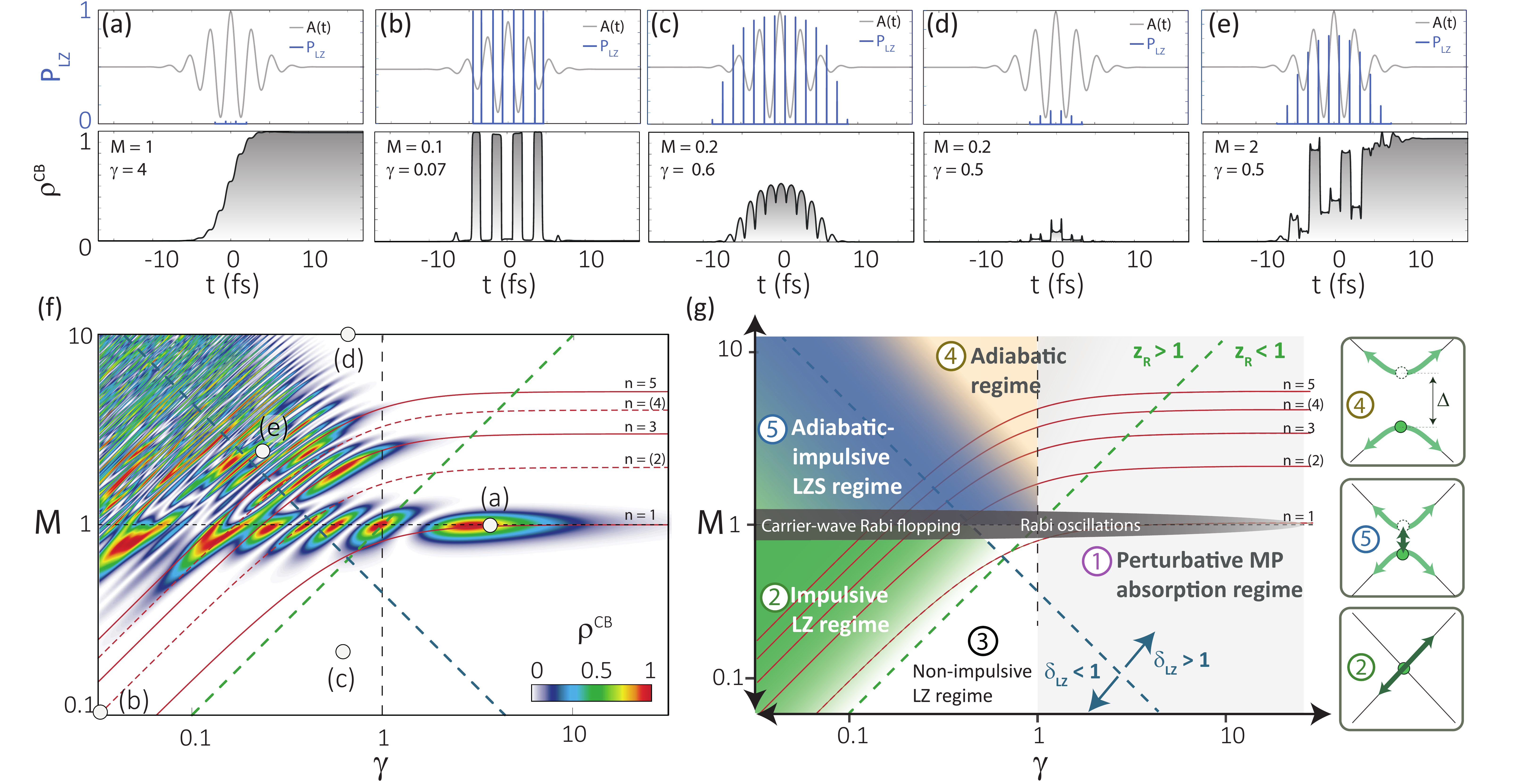}
			\caption{\label{Figure2} \textbf{Regimes of a light field-driven two-level system.} 
				\textbf{(a--e)} Temporal evolution of the conduction band population $\rho^\text{CB}$ for a Gaussian pulse ($\tau_\text{P}$ = 5\,fs, $\phi_\text{CEP} = \pi/2$) with electrons starting from $k_0 = 0$ from our TDSE model. The top panels show the normalized vector potential $A(t)$ (grey) and $P_\text{LZ}$ (blue) for an LZ transition at $k_0$. The laser and system parameters are identified by $\gamma$ and $M$, given in each legend. \textbf{(f)} Map of the residual conduction band population. The filled spheres relate to the panels (a--e). See text for details. \textbf{(g)} same as (f), with various regimes indicated: (1) Perturbative MP absorption regime, (2) impulsive LZ regime, (3) non-impulsive LZ regime, (4) adiabatic regime and (5) adiabatic-impulsive regime.	Whereas for $\gamma > 1$ only perturbative, resonant one-photon ($n = 1$) absorption at $M = 1$ is found, for $\gamma < 1$ off-resonant excitation occurs. In particular for $P_\text{LZ} \approx 0.5$, $\delta_\text{LZ} \approx \ln(2)/(2\pi)$ and $z_\mathcal{R} > 1$, defining the adiabatic-impulsive regime (5), most off-resonant excitation is found. The red lines represent odd (solid) and even (dashed) resonances for the light matter interaction, based on Eq.\,\eqref{Eq. analytic ponderomotive}. The boxes relate to three regimes as indicated by the number. Note the dashed lines showing the crossing areas for the various adiabaticity parameters.} 
		\end{center}
	\end{figure*}
	
	In his seminal paper, Keldysh introduced the adiabaticity parameter $\gamma \equiv \omega\sqrt{m\Delta}/(eE_0)$ for electrons in a TLS \cite{Keldysh1965}. 
	Here, $\omega$ is the driving frequency of the light, $E_0$ its peak electric field strength, $m$ the effective electron mass, $\Delta$ the minimal energy spacing between two bands and $e$ the electron charge. Assuming monochromatic light and excitation at $\Delta$, we can express the Keldysh parameter as
	\begin{align}
	\gamma = \frac{\Delta}{2\hbar\Omega_\text{R}}
	\label{eq. gamma}
	\end{align}
	with $\Omega_\text{R} = v_\text{F}eE_0/(\hbar \omega)$ the Rabi frequency and $v_\text{F}$ the Fermi velocity, see Fig.\,\ref{Figure1} and SI for a detailed derivation. When $2\hbar\Omega_\text{R}$ exceeds $\Delta$, $\gamma$ becomes smaller than 1 and the light-matter interaction enters the well-known strong-field regime. \\
	
	In case of resonant excitation, $\Delta = \hbar\omega$, we obtain the resonant adiabaticity parameter from Eq.\,\eqref{eq. gamma} \cite{Reiss1980, Wegener2005, Kruchinin2018} as
	\begin{align}
	z_\mathcal{R} = \frac{2\Omega_\text{R}}{\omega}.
	\label{Eq. zR}
	\end{align}
	Here, $2\Omega_\text{R}$ can be interpreted as the inverse of the transition time $\tau_t$ ($2\Omega_\text{R} = (2\pi)/\tau_t$) from the valence to the conduction band. Figure \ref{Figure1} shows schematically the competition of Rabi oscillations with $\Omega_\text{R}$ (blue arrow) and intraband motion, driven with $\omega$ (thick green arrows). When $\tau_t$ becomes shorter than the driving period of the light, $z_\mathcal{R}$ becomes larger than 1. 
	In that regime, colloquially speaking, the electron has enough time to undergo a transition from one to the other band within an optical cycle.\\ 
	
	We will continue with further insightful parameters later but now first model the light-matter interaction by solving the time-dependent Schrödinger equation numerically. The resulting excitation maps will help us to identify the various regimes. We consider a TLS with two energy levels $\pm\Delta/2$ and a time-dependent perturbation $\pm \alpha(t)/2$, representing an avoided crossing \cite{Landau1932, Zener1932, Kayanuma1997, Ashhab2007, Shevchenko2010}. The Hamiltonian of this system reads:
	\begin{align}
	\label{Ham}
	\hat{{\cal H}}(t) = -\frac{\alpha(t)}{2} \hat{\sigma}_\text{x} - \frac{\Delta}{2} \hat{\sigma}_\text{z},
	\end{align}
	with $\hat{\sigma}_\text{x, z}$ the Pauli matrices. The eigenenergies are $\varepsilon_{\pm}(t) \equiv \pm\frac{1}{2}\varepsilon(t)$, where $\varepsilon(t)=\sqrt{\Delta^2 + \alpha(t)^2}$ is the time-dependent energy difference between the two levels. By taking $k(t)$ as the time-dependent wave number of an electron, and $\pm v_\text{F}$ as the slopes of the two crossing levels, Eq.~\eqref{Ham} with $\alpha(t) = 2\hbar v_\text{F} {k}(t)$ represents a solid state band structure with $\varepsilon_{\pm}(k)$ the conduction band (+) and valence band energy (--) (and $\Delta$ the band gap). We assume $v_\text{F} = 1\,$nm/fs. The change of the electron wave number due to the electric field is described by the Bloch acceleration theorem $\dot{k}(t) = -(e/\hbar)E(t)$ \cite{Bloch1929, Kelardeh2015, Chizhova2017, Higuchi2017}. We note that when the electric field is weak, the electron dynamics is well described with a trivial TLS of fixed levels $\varepsilon_{\pm}$. However, when the electric field becomes large, different k-values become coupled and the electron undergoes intraband motion, probing the band structure. For even larger electric field strengths (i.e., $\gamma < 0.01$), relativistic effects may become important, which are beyond the presented categorization (see SI).  \\
	In the simulations we apply a linearly polarized vector potential 
	\begin{align}
	A(t)= -E_0/\omega\exp{(-2\ln{2}(t/\tau_p)^2)}\sin{(\omega t+\phi_\text{CEP})} ,
	\end{align}
	associated with an electric field $E(t)=-\dot{A}(t)$, to model the temporal evolution of the conduction band population. By defining the vector potential we satisfy $A(-\infty)=A(\infty)$ and thus omit dc components in the electric field. A pulse duration of $\tau_p=5$\,fs is chosen with central photon energy $\hbar\omega=1.55$\,eV and a carrier-envelope phase $\phi_\text{CEP}=\pi/2$. For simplicity, we first consider electrons with an initial wave number of $k_0 = 0$. Note that while different pulse durations, dephasing and dispersion effects may change the population distribution, the presented categorization remains valid (see SI).\\
	
	Figures \ref{Figure2}\,(a--e) show in the lower panels the conduction band population $\rho^{\text{CB}}$ as a function of time for various regimes, discussed in detail in what follows. In Fig.\,\ref{Figure2}\,(f), we show the residual conduction band population parameterized by $\gamma$ and the multi-photon parameter $M = \Delta/(\hbar\omega)$, measuring the band gap in units of the photon energy. Based on $\gamma$ and $z_\mathcal{R}$ we can now identify and categorize various regimes, which show an entirely different temporal evolution of the conduction band population [Fig.\,\ref{Figure2}\,(a--e)].\\
	\textbf{(1) Perturbative multi-photon absorption regime:} When the electric field is weak ($\gamma > 1$), population is found at $M \approx 1$, reflecting resonant excitation [Fig.\,\ref{Figure2}\,(f, g)]. Here, the population $\rho^{\text{CB}}$ rises during the laser pulse gradually [Fig.\,\ref{Figure2}\,(a)]. The red lines, horizontal for $\gamma > 1$, represent the resonance condition: $M$ is integer. The population width along $M$ reflects the spectral width of the laser pulse. Increasing the electric field strength (decreasing $\gamma$) results in an increase of $\Omega_\text{R} \propto E_0$, hence Rabi oscillations become visible.\\ 
	\textbf{Transition to the strong-field regime (2)-(5)}:
	Around $\gamma \approx 1$, the interaction strength $2\hbar\Omega_\text{R}$ is of the order of the band gap [Eq.\,\eqref{eq. gamma}] and light field-driven intraband motion cannot be neglected anymore. Hence, intraband motion leads to a variation of $\varepsilon(t)$, which in turn results to a rotation of the features in Fig.\,\ref{Figure2}\,(f). For $\gamma < 1$, carrier-wave Rabi flopping occurs and the horizontal lines given by multiples of the photon energy fail as a valid quantity to specify resonant absorption, reflecting the AC Stark effect \cite{Wismer2016}. To extend the resonance condition to this field-driven regime, we calculate the dynamical phase
	\begin{align}
	\phi = \frac{1}{\hbar}\int \varepsilon(t) \text{d}t.
	\label{equ. pulse area theorem}
	\end{align}
	When a phase of $\phi=2\pi$ is accumulated within an optical cycle of the laser pulse, population at the next higher multi-photon resonance ($n$) is found \cite{Wismer2016}.
	For monochromatic excitation, we analytically obtain the resonance condition
	\begin{align}
	\label{Eq. analytic ponderomotive}
	M = \frac{n\pi}{2\mathfrak{e}_l(-\gamma^{-2})}.
	\end{align}
	Here, $\mathfrak{e}_l$ denotes the complete elliptic integral of the second kind and $n$ the order of the resonance (see SI). We see that when $\gamma$ decreases, the photon resonances $n$ no longer match integer multiples of $M$ but are shifted towards smaller $M$. Equation \eqref{Eq. analytic ponderomotive} is plotted as red solid lines in Fig.\,\ref{Figure2}\,(f), for various $n$, perfectly matching the conduction band population obtained from the TDSE simulation. When $\gamma \gg 1$, the elliptic integral becomes $\pi/2$ and the photon resonances are found at integer multiples of the band gap, i.e., $M = n$, as expected for weak fields. In contrast, at $\gamma \approx 0.4$ the intraband motion becomes so strong that $n = 2M$ and, thus, the \textit{two}-photon resonance is found at $M = 1$. We note that due to the inversion symmetry of the TLS at $k_0 = 0$ population only arises for odd photon orders [see dashed lines for even orders in Fig.\,\ref{Figure2}\,(f)].\\ 
	In the field-driven regime, i.e., $\gamma < 1$ it is helpful to describe the electron dynamics within the Landau-Zener (LZ) formalism, see SI and \cite{Rubbmark1981, Shapiro1999, Ashhab2007, Shevchenko2010, Chizhova2016}. At time $t'$, when the electron reaches the minimal separation of valence and conduction band, the transition from one to the other band takes the form of an LZ transition. The transition probability is approximated by the famous Landau-Zener formula
	\begin{align}
	\label{eq. Landau-Zener probability}
	P_\text{LZ} = \exp\left(-2\pi\delta_\text{LZ}\right),
	\end{align}
	with 
	\begin{align}
	\delta_\text{LZ} \equiv \frac{1}{\hbar\alpha(t')} \left(\frac{\Delta}{2}\right)^2
	\label{Eq. LZ parameter}
	\end{align}
	the Landau-Zener adiabaticity parameter \cite{Landau1932, Zener1932, Shevchenko2010}.\\
	Within the field-driven regime we now find four categories of electron dynamics.\\
	\textbf{(2) Impulsive Landau-Zener regime:} 
	When $\delta_\text{LZ} \ll 1$ and $z_\mathcal{R} > 1$ the electron undergoes a sequence of fast LZ transitions with probability $P_{\rm LZ}$ close to unity [Fig.\,\ref{Figure2}\,(b)]. Within one optical cycle the electron experiences a transition from the valence to the conduction band and back to the valence band. Because of $z_\mathcal{R} > 1$, the regime is called impulsive. After the laser pulse, the electron ends up in the initial band, hence no excitation is found [Fig.\,\ref{Figure2}\,(f, g)]. \\
	\textbf{(3) Non-impulsive Landau-Zener regime:} 
	When the transition time $\tau_t$ is longer than the optical period (i.e., $z_\mathcal{R} < 1$), the LZ transition can no longer be considered impulsive. Figure \ref{Figure2}\,(c) shows that even when the LZ probability is large (blue lines in top panel) there is not enough time for an efficient electron excitation. Also here, the conduction band is not populated after the laser pulse is gone, albeit for very different reasons than in (2) [Fig.\,\ref{Figure2}\,(f)].\\
	\textbf{(4) Adiabatic regime:}
	When $\delta_\text{LZ} \gg 1$, the probability for an electron to undergo an LZ transition approaches zero and, thus, the electron undergoes pure intraband motion [Fig.\,\ref{Figure2}\,(d)]. Yet, we note that the intraband motion during the laser pulse is a well-investigated source of intraband HHG \cite{Vampa2014, Luu2015}.\\
	\textbf{(5) Adiabatic-impulsive Landau-Zener-Stückelberg regime:} When $\delta_\text{LZ} \approx \ln(2)/(2\pi)$, $P_\text{LZ} \approx 0.5$. Hence after one LZ transition event, the electron wave function is equally split  into valence and conduction band, so a part of the electron wave function undergoes an LZ transition, while the remainder stays adiabatically in the valence band [Fig.\,\ref{Figure2}\,(e)]. Due to the oscillatory nature of the driving, this happens periodically with every half cycle of the laser pulse. Interference of the electronic wave function components, each with a different accumulated phase $\phi$ [Eq.\,\eqref{equ. pulse area theorem}], determines the conduction band excitation probability. We observe that, intriguingly, the net excitation probability is highest in this regime. We note that repeated coherent LZ transitions is called Landau-Zener-Stückelberg interference \cite{Landau1932, Zener1932, Kayanuma1997, Ashhab2007, Shevchenko2010}.\\ 
	\begin{figure}[h!]
		\begin{center}
			\includegraphics[width=8cm]{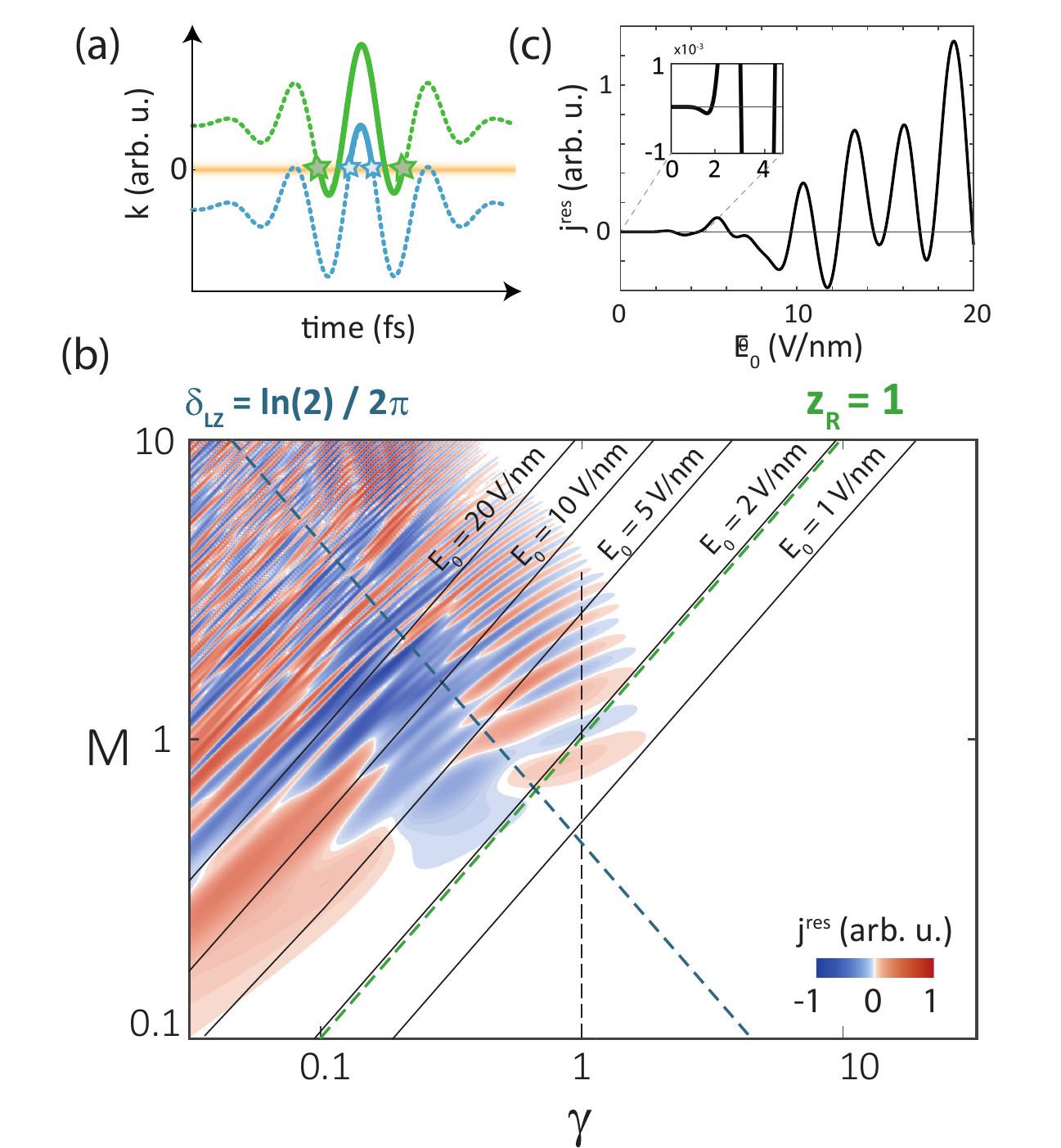}
			\caption{\label{Figure3}  
				\textbf{(a)} Temporal evolution of the electron trajectory for an electron with initially positive (green line) and negative (blue line) wave number. The orange line indicates the region of high LZ transition probability. For a field waveform with broken time inversion symmetry, the two electron trajectories can experience a different number of LZ transitions and/or a different total quantum mechanical phase evolution, in particular when, the length of the trajectory (green and blue) is different. 
				\textbf{(b)} Residual electric current as function of $\gamma$ and $M$ calculated with the TDSE model. The maximal residual current is found in the adiabatic-impulsive LZS regime. The black lines indicate iso-lines for constant $E_0$. \textbf{(c)} Integrated current along lines with constant $E_0$, as a function of $E_0$. The oscillatory nature of $j^\text{res}(E_0)$ is a result of the wave number-dependent quantum mechanical phase evolution.} 
		\end{center}
	\end{figure}

	So far, we have discussed electrons starting from $k_0~=~0$ only. Electrons with $k_0 > 0$ and $k_0 < 0$ can experience different dynamics, in particular, when a pulse with a broken time inversion symmetry ($E(-t) \neq E(t)$) is applied, e.g., one with $\phi_{\rm CEP} = \pm\pi/2$ \cite{Franco2008}. Figure\,\ref{Figure3}\,(a) shows schematically the electron trajectory in reciprocal space for an electron starting at $k_0 > 0$ (green line) and $k_0 < 0$ (blue line). For $\phi_{\rm CEP} = \pm\pi/2$, the number of LZ transitions and their temporal spacing between an electron starting at initially positive or negative wave number may differ, resulting in asymmetric residual population. As a consequence, a nonzero residual ballistic current
	\begin{align}
	\label{Eq. current from cond pop}
	j^\text{res} = g_se\sum_{m=\text{CB},\text{VB}}\int_{-\infty}^{\infty} v^{(m)}(k(t)) \rho^{(m)}(k_0,t) \frac{\text{d}k}{2\pi},
	\end{align}
	with $v^{(m)}(k) = \hbar^{-1}\frac{\partial \varepsilon_\pm(k)}{\partial k}$ is generated. The factor $g_s = 2$ accounts for two kinds of spins.\\ 	
	In Fig.\,\ref{Figure3}\,(b) we show the map of the residual current density, taking all initial $k$-values into account. With the help of the above, we can now understand the intricate pattern of the current map. Residual current is mainly found in the adiabatic-impulsive LZS regime. Increasing the electric field strength, i.e., decreasing $\gamma$, starting at $\gamma \approx 1$, results in an increase of the accumulated dynamical phase, causing more and more current reversals as a function of $E_0$. Iso-lines for constant electric field strengths $E_0 = 1 ... 20\,$V/nm are drawn as solid black lines. Around $E_0 = 1$\,V/nm almost no residual current is obtained. Increasing the electric field strength towards 2\,V/nm results in nonzero $j^\text{res}$, at around 2\,V/nm. This is shown and continued in the integrated current, along constant field strength in Fig.\,\ref{Figure3}\,(c). The oscillatory nature of the current as a function of the electric field strength reflects Landau-Zener-Stückelberg interference with varying accumulated phase, as also discussed above.\\

	The current map obtained here numerically also matches an important previous result: In \cite{Higuchi2017} it was shown that the CEP-dependent photocurrent switches sign at a field strength of 1.8\,V/nm, perfectly coinciding with the first current reversal shown in Fig.\,\ref{Figure3}\,(c). Intriguingly, the adiabaticity parameters here are $\gamma < 1 $ and $z_\mathcal{R} > 1$, indicating that the adiabatic-impulsive LZS regime has been entered in experiment.\\	
	{Our TLS numerical results are based on the example of a narrow-band gap band structure to provide a connection to experimental conditions as given by strong-field physics in solids. However, the discussion is applicable to any avoided level crossing by appropriate transformation of parameters. For example, in wide-band gap systems, the Bloch frequency rather than the Rabi frequency is used to define $z_\mathcal{R}$ \cite{Kruchinin2018}. Here, Wannier-Stark (WS) localization and Bloch oscillations are the dominating mechanism for LZ transitions \cite{Schiffrin2013}. \\
		In cavity QED, the characteristic system coupling $g$ is compared with $\omega$ to categorize excitation regimes \cite{FriskKockum2019, FornDiaz2019}. 
		In particular, in the ultrastrong coupling regime ($g/\omega > 0.1$) and the deep strong coupling regime ($g/\omega > 1$), the rotating wave approximation breaks down and new quantum phenomena emerge \cite{Bayer2017}. In full analogy to $g/\omega$, we distinguish between perturbative Rabi physics and the non-impulsive LZ regime for $z_\mathcal{R} < 1$ and the impulsive strong-field regimes for $z_\mathcal{R} > 1$.}
	
	In summary, we have presented five different regimes of light-matter interaction in a two-level system, categorized by adiabaticity parameters. This way, we can understand the intricate dynamics of each regime by delimiting it from but also connecting it to its neighboring regimes. Although we have focused on electrons in a solid state band structure, the here discussed dynamics and categorization into different regimes is well applicable to a large variety of two-level systems, which may include or give rise to nonadiabatic multielectron dynamics \cite{Shapiro1999, Lezius2002}, conical intersections \cite{Worner2011, Kling2013}, Wannier-Stark localization \cite{Wannier1960a, Schiffrin2013}{, cavity QED \cite{FriskKockum2019,FornDiaz2019},} and the Kibble-Zurek mechanism \cite{Damski2005}, for which the time-dependent perturbation $\alpha(t)$ needs to be adapted appropriately, while the rest of the model is unchanged. Specifically, 2D materials can be well approximated as two-level systems \cite{Higuchi2017, OliaeiMotlagh2018, OliaeiMotlagh2020}. Furthermore, our work is directly relevant to quantum information processing \cite{Ashhab2007, Berns2006, Buluta2011}, where coherent electronics based on LZS interference has been recently implemented \cite{Otxoa2019}. We expect that the presented categorization will help to understand fundamental yet complex light-matter interaction results on a new level across a large variety of systems.

	\begin{widetext}
		\begin{center}
			\large \textbf{Appendix}\\ 
		\end{center}
		\vspace{1cm}
	\end{widetext}
	
	
	\subsection{General form of a two-level system}
\begin{figure}[h!]
		\begin{center}
			\includegraphics[width=5.5cm]{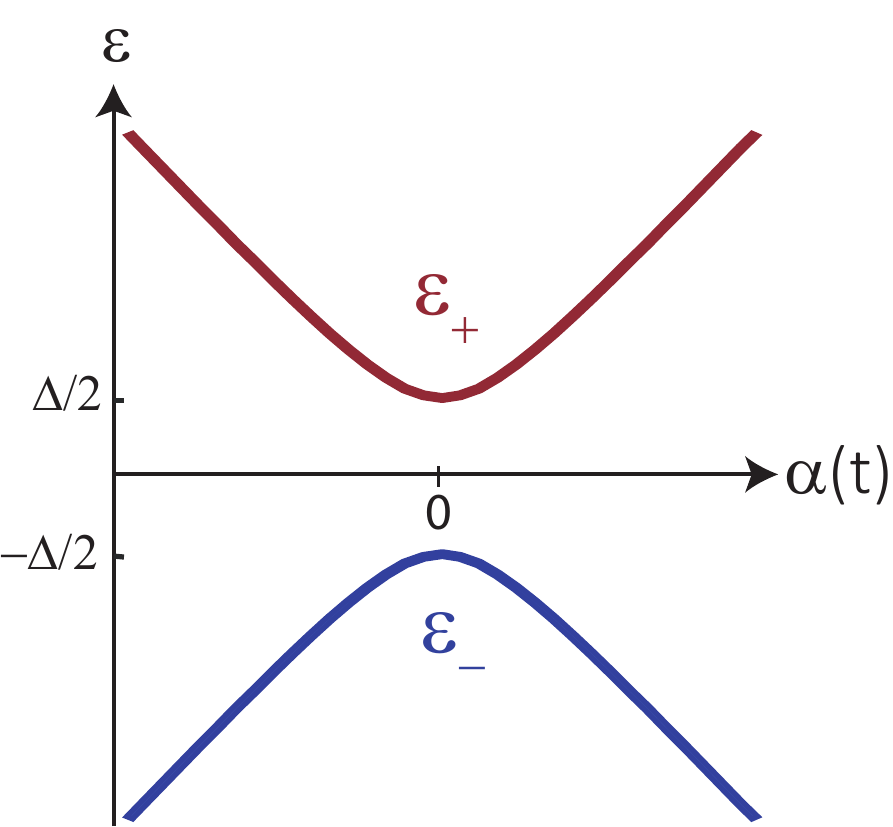}
			\caption{\label{S1} \textbf{Two-level system.} Energy $\varepsilon$ versus time-dependent energy bias $\alpha$. The blue and red solid lines represents the adiabatic energy basis of the two-level system with two instantaneous eigenstates  $\varepsilon_{+}$ and $\varepsilon_{-}$. For $\alpha = 0$, the energy difference $\varepsilon(t)$ between the two eigenstates is given by $\Delta$.} 
		\end{center}
	\end{figure}

	We consider a two-level system (TLS) with  two constant energy levels $\pm \Delta/2$ and a time-dependent energy perturbation $\alpha(t)/2$ acting on them, i.e., time-dependent coupling between energy levels. Such a system is described by the Hamiltonian 
	\begin{equation}
	\begin{split}
	\hat{\cal H}(t) &= \hat{\cal H}_0 + \hat{\cal H}_\text{int}(t) \\
	& = -\frac{1}{2}
	\begin{pmatrix}
	\Delta &  \alpha(t)\\
	\alpha(t) & -\Delta 
	\end{pmatrix}.
	\end{split}
	\label{Eq. two-level ham1}
	\end{equation}
	Here, $\hat{\cal H}_0 = -\frac{\Delta}{2}\hat{\sigma}_\text{z}$ represents the bare two-level system with two eigenvalues $\pm\Delta/2$ and $\hat{\cal H}_\text{int}(t) = -\frac{\alpha(t)}{2} \hat{\sigma}_\text{x}$ the interaction Hamiltonian, describing how the system evolves under a time-dependent perturbation. $\hat{\sigma}_\text{x}$ and $\hat{\sigma}_\text{z}$ are Pauli matrices. By diagonalizing $\hat{\cal H}(t)$, one obtains the instantaneous eigenvalues
	\begin{align}
	\label{Eq. eigenstate}
	\varepsilon(t)_\pm \equiv \pm \frac{1}{2} \sqrt{\Delta^2 + \alpha(t)^2} = \pm \frac{1}{2} \varepsilon(t). 
	\end{align}
	Hereby, 
	\begin{align}
	\label{Eq. omega(alpha)}
	\varepsilon(t) = \sqrt{\Delta^2 + \alpha(t)^2}
	\end{align}
	is the time-dependent energy difference between the two energy states $\varepsilon_{+}$ and $\varepsilon_{-}$, as depicted in Fig. \ref{S1} .\\
	
	\subsection{Solid-state two-level system}
	\label{chap. theory sec. Formalism}
	One example of a TLS system can be found in the dynamics of an electron driven in a solid by an external field. 
	By taking the Bloch acceleration theorem 
	\begin{align}
	\label{Eq. trajectory}
	k(t) = k_0 - \frac{e}{\hbar}\int E(t) \text{d}t,
	\end{align}
	as the time-dependent electron wave number, where $A(t)$ is the vector potential and $k_0$ the initial electron wave number, and 
	\begin{align}
	\label{Eq. alpha}
	\alpha(t) = 2 \hbar v_\text{F} k(t)
	\end{align}
	as the time-dependent energy bias, with $v_\text{F}$ the Fermi velocity, we can rewrite Eq.\,\eqref{Eq. two-level ham1} to
	\begin{align}
	\label{Eq. two-level ham2}
	\hat{\cal H}(t) = 
	\left(
	\begin{array}{cc}
	-\frac{\Delta}{2} & \hbar v_\text{F} k(t) \\
	\hbar v_\text{F} k(t) & \frac{\Delta}{2}
	\end{array}
	\right). 
	\end{align}
	For $k_0 = 0$ and a periodic oscillating electric field $E(t) = E_0\sin(\omega t)$, Eq.\,\eqref{Eq. two-level ham2} has the form as that of a driven two-level Rabi system with
	\begin{align}
	\label{eq. Rabi Frequency}
	\Omega_\text{R} = \frac{v_\text{F}eE_0}{\hbar \omega}.
	\end{align}
	the Rabi frequency. $\omega$ is the angular frequency and $E_0$ the peak electric field strength.\\
	
	To obtain the temporal evolution of an electron in the two-level system, we assume that the electron dynamics is coherent and can thus be described by the time-dependent Schrödinger equation (TDSE)
	\begin{align}
	i\hbar \frac{\partial\Psi(k_0,t)}{\partial t} = \hat{\cal{H}}(t) \Psi(k_0,t)
	\label{Eq. TDSE}
	\end{align}
	with $\Psi(k_0,t) = \sum_{m=\text{CB},\text{VB}} \beta^{(m)}(k_0,t)\Phi(k_0)$ the electron wave function with  $\Phi$ the Bloch states of the field-free system and $\beta^{(m)}$ the expansion coefficients of Eq.\,\eqref{Eq. TDSE} \cite{OliaeiMotlagh2020}.
	\subsection{Keldysh adiabaticity parameter}
	The general form of the Keldysh parameter for a two-level system reads
	\begin{align}
	\gamma = \frac{\omega\sqrt{m\Delta}}{eE_0}.
	\end{align}
	with $m$ the effective electron mass \cite{Keldysh1965}.
	Using Eqs.\,\eqref{Eq. omega(alpha)} and \eqref{Eq. alpha} we obtain
	\begin{align}
	m = \hbar^2 \left[\frac{\partial^2\varepsilon(k)}{\partial k^2}\Big|_{k_0=0}\right]^{-1} = \frac{\Delta}{4v_\text{F}^2}.
	\label{Eq. mass}
	\end{align}
	Applying the Rabi frequency [Eq.\,\eqref{eq. Rabi Frequency}], we rewrite the Keldysh adiabaticity parameter for an oscillating electric field
	\begin{align}
	\gamma = \frac{\omega \Delta}{2v_\text{F} e E_0} = \frac{\Delta}{2 \hbar \Omega_\text{R}}.
	\end{align}
	Here, the Keldysh adiabaticity parameter is given as the characteristic interaction strength $2\hbar\Omega_\text{R}$ and the band gap $\Delta$. Note that in case of resonant excitation $\hbar\omega = \Delta$ and $\omega \ll \Omega_\text{R}$, this regime is known as carrier-wave Rabi flopping.  
	\subsection{Landau-Zener transition}
	Following the framework of Landau-Zener (LZ) transitions, it is useful to work with a Hamiltonian $\hat{\cal H'}(t)$ with time-dependent diagonal components, which can be achieved by applying the unitary transformation
	\begin{align}
	\hat{U} = \exp\left(i\frac{\pi}{4}\hat{\sigma}_\text{y}\right).
	\end{align}
	We obtain 
	\begin{equation}
	\begin{split}
	\hat{\cal H'}(t) &= \hat{U}\hat{{\cal H}}(t) \hat{U}^{-1}\\
	&= -\frac{\Delta}{2}\hat{\sigma}_\text{x}-\frac{\alpha}{2}\hat{\sigma}_\text{z}.
	\end{split}
	\label{Eq. two-level ham4}
	\end{equation}
	In the vicinity of an LZ transition we linearize $\alpha(t)$, i.e., $\frac{\cos(\omega t)}{\omega} \approx t$ and obtain $\alpha(t) \approx \alpha_0 t$, with $\alpha_0 = 2 v_\text{F} e E_0$. Now, $\hat{\cal H'}(t)$ reads
	\begin{align}
	\hat{\cal H'}(t) &= -\frac{1}{2}\left(\begin{array}{cc}
	\alpha_0 t & \Delta \\
	\Delta & -\alpha_0 t
	\end{array}
	\right).
	\label{Eq. two-level ham3}
	\end{align}
	This Hamiltonian has the form as a Hamiltonian for an avoided crossing model with the Landau-Zener transition probability
	\begin{align}
	P_\text{LZ} = \exp\left(-2\pi\delta_\text{LZ}\right),
	\end{align}
	with $\delta_\text{LZ} = \Delta^2/(4\hbar\alpha)$, the Landau-Zener adiabaticity parameter \cite{Shevchenko2010}. For $\delta_\text{LZ} \ll 1$, $P_\text{LZ} \rightarrow 1$ and an interband transition becomes likely.  
	\subsection{Analytic resonance condition}
	When an electric field is applied to solids, the wave number $k(t)$ of an electron changes according to Eq.\,\eqref{Eq. trajectory}. Within one optical cycle, the resulting intraband motion leads to the accumulation of a dynamical phase
	\begin{align}
	\phi = \frac{1}{\hbar} \int^{T_0}_0\varepsilon(t) \text{d}t
	\label{equ. pulse area theorem}
	\end{align}
	with $T_0 = 2\pi/\omega$. We assume a periodically oscillating electric field to obtain an analytic expression for the resonance condition under the presence of intraband motion. Using the Keldysh parameter, we rewrite Eq.\,\eqref{Eq. omega(alpha)}:
	\begin{equation}
	\begin{split}
	\varepsilon(t) &= \sqrt{\Delta^2 + \left(2v_\text{F} e \frac{E_0}{\omega} \cos(\omega t) \right)^2} \\
	&= \Delta \sqrt{1+ \left(\frac{\cos(\omega t)}{\gamma}\right)^2}.
	\end{split}
	\end{equation}
	Taking $\omega t = t'$ and $t'' = t' -\frac{\pi}{2}$ we write
	\begin{equation}
	\begin{split}
	\phi &= \frac{2\Delta}{\hbar \omega} \cdot \int_0^{\pi} \sqrt{1+ \left(\gamma^{-1}\cos(t')\right)^2} \text{d}t'\\
	&= \frac{4\Delta}{\hbar \omega} \cdot \int_0^{\frac{\pi}{2}} \sqrt{1+ \left(\gamma^{-1}\sin(t')\right)^2} \text{d}t''.
	\end{split}
	\end{equation}
	Now we apply the complete second order elliptic integral 
	\begin{align}
	\phi = \frac{4\Delta}{\hbar \omega}  \mathfrak{e}_l(\gamma^{-2}).
	\end{align}
	When a phase of $\phi=2\pi$ is accumulated, population at the next higher multi-photon resonance ($n$) is found. Thus, 
	\begin{align}
	\frac{4\Delta}{\hbar \omega}  \mathfrak{e}_l(\gamma^{-2}) \stackrel{!}{=} 2\pi\cdot n
	\end{align}
	needs to be fulfilled and the 
	the condition for analytical resonance under the presence of intraband motion reads
	\begin{align}
	M = \frac{\pi n}{2 \mathfrak{e}_l(\gamma^{-2})}.
	\end{align}
	For $\gamma \gg 1$, the elliptic integral becomes $\frac{\pi}{2}$ and the photon resonances are found at multiples of the bandgap ($M=n$).
	\begin{figure*}[t!]
	\begin{center}
		\includegraphics[width=17cm]{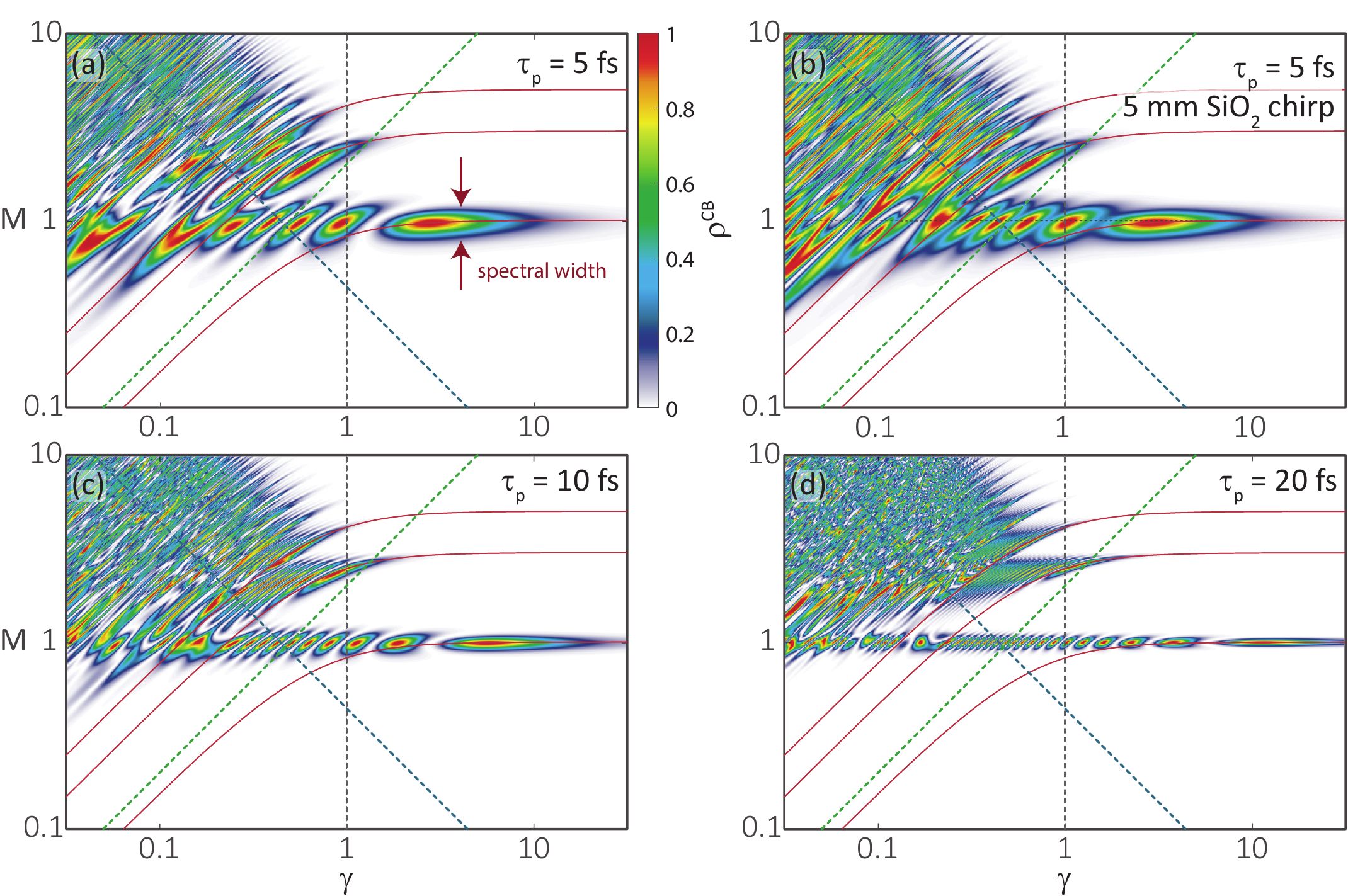}
		\caption{\label{S2} \textbf{Different pulse durations and dispersion.} Map of the residual conduction band population for three different bandwidth-limited pulse durations (5\,fs (a), 10\,fs (c), 20\,fs (d)) and for a 5\,fs pulse with dispersion of 5\,mm SiO$_2$ (b). Whereas for $\gamma > 1$ only perturbative, resonant one-photon ($n = 1$) absorption at $M = 1$ is found, for $\gamma < 1$ off-resonant excitation occurs. In particular for $P_\text{LZ} \approx 0.5$, $\delta_\text{LZ} \approx \ln(2)/(2\pi)$ and $z_\mathcal{R} > 1$, defining the adiabatic-impulsive regime (5), most off-resonant excitation is found. The red lines represent odd resonances for the light matter interaction. The dashed green line represents $z_\mathcal{R} = 1$ and the dashed blue line $P_\text{LZ} \approx 0.5$. See main text for detailed discussion.} 
	\end{center}
\end{figure*}
	\subsection{Genericity under pulse distortion and dephasing}
\begin{figure}
		\begin{center}
			\includegraphics[width=9cm]{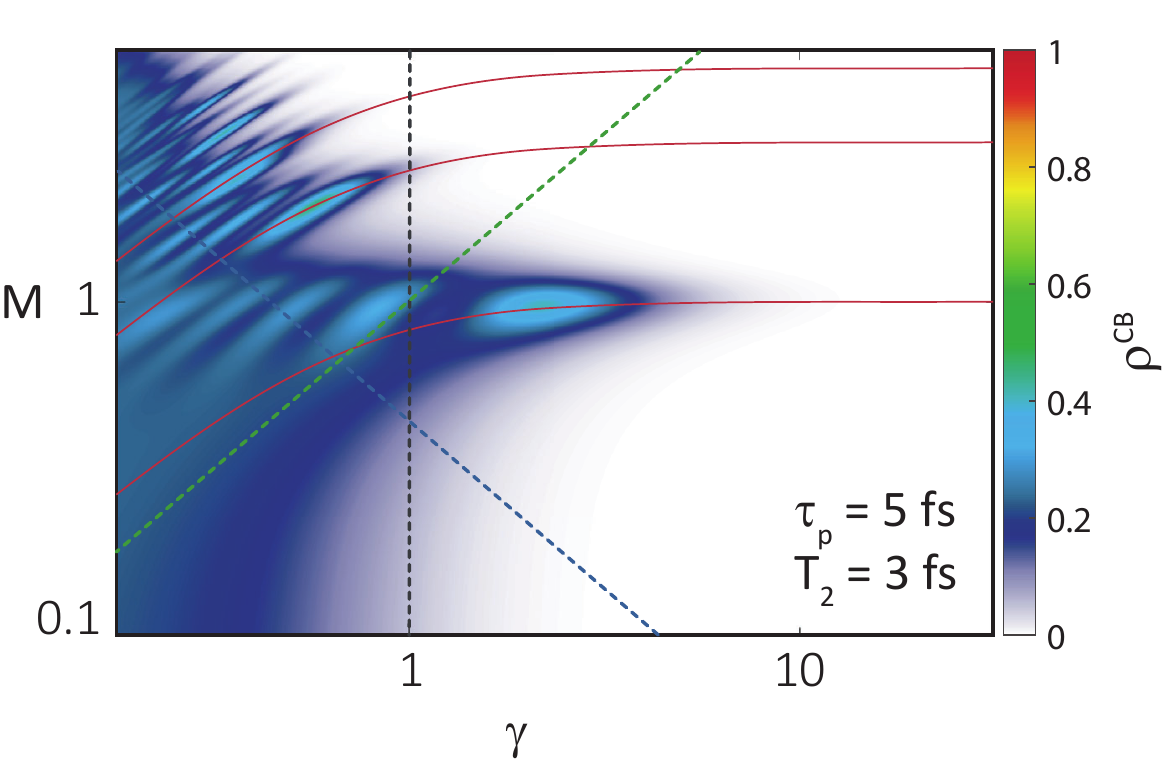}
			\caption{\label{S3} \textbf{Dephasing of population.}  Map of the residual conduction band population assuming a dephasing time constant of $T_2=3$\,fs and a pulse duration of 5\,fs. The red lines represent odd resonances for the light matter interaction. The dashed green line represents $z_\mathcal{R} = 1$ and the dashed blue line $P_\text{LZ} \approx 0.5$. See main text for detailed discussion.} 
		\end{center}
	\end{figure}

	The categorization of excitation regimes in a strongly-driven two-level system is general and remains valid for different pulse durations, dispersion and dephasing, i.e., $z_\mathcal{R}$, $\delta_\text{LZ}$ and $\gamma$ depends on the driving frequency, peak electric field strength and the energy spacing between the two bands. To prove this, we show in Fig.\,\ref{S2} the residual conduction band population for three different bandwidth-limited pulse durations (Fig.\,\ref{S2}\,a: 5\,fs, b: 10\,fs, c: 20\,fs) centered at a photon energy of $\hbar\omega = 1.55$\,eV. Increasing the pulse duration results in a decrease in the spectral width of the photon orders and an increase of Rabi cycles as a function of $\gamma$, i.e., more Rabi cycles can be performed during the laser pulse. In the field-driven regimes, increasing the number of optical cycles results in more subsequent LZ-transitions and a different accumulated dynamical phase. In the adiabatic-impulsive LZS regime, both lead to strong mixing of states resulting in finer features in the residual conduction band population. Whereas the exact value of $\rho^\text{CB}$ depends on the pulse shape, the categorization into different excitation regimes is determined by peak electric field strength.\\
	Similarly, introducing dispersion does not affect the categorization. In Fig.\,\ref{S2}\,b we show the residual conduction band population of the same pulse applied in Fig.\,\ref{S2}\,a, but stretched by 5\,mm SiO$_2$ (GDD: 180.8\,fs$^2$, TOD: 137.3\,fs$^3$).  The visibility of the Rabi-oscillations is decreased since the spectral components are delayed with respect to each other. In the field-driven regimes, all features are preserved while again, the exact value of conduction band population is changed due to different accumulation of dynamical phase.\\
	Figure\,\ref{S3} shows the residual conduction band population for a pulse duration of 5\,fs including a dephasing time constant $T_2 = 3$\,fs \cite{Wismer2016}. Decreasing the dephasing time results in an increase of $\rho^\text{CB}$ in the adiabatic regime, i.e., transiently populated carriers remain after the laser pulse and the residual population in the adiabatic-impulsive LZS regime becomes smeared out. LZS interference becomes less important due to decoherence and the visibility of features decreases consequently. We note that the dephasing time constant $T_2$ can be understood as decay of quantum coherence in a coherently driven system.  
	
	\subsection{Relativistic regime}
	\begin{figure}
		\begin{center}
			\includegraphics[width=7cm]{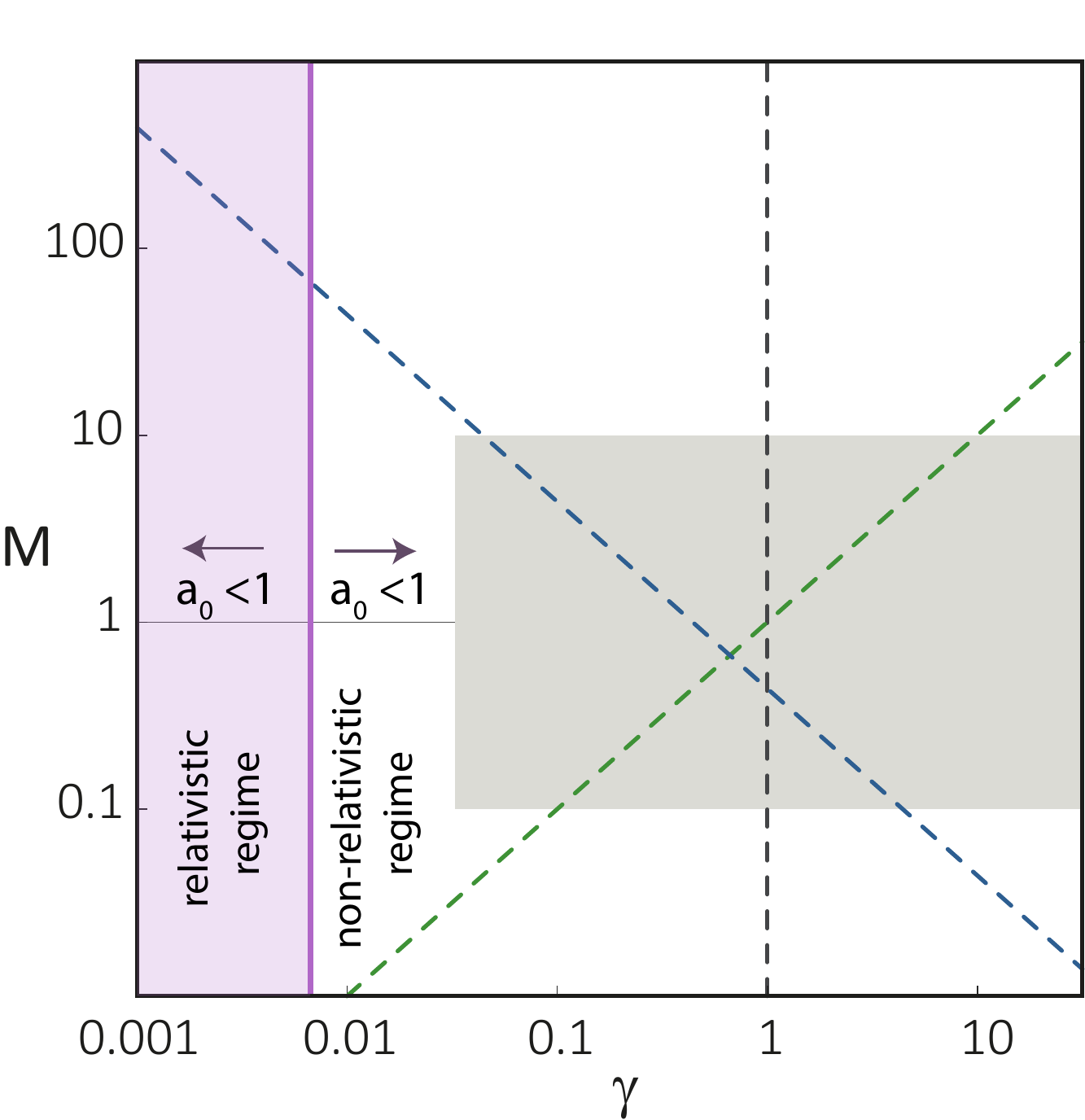}
			\caption{\label{S5} \textbf{Relativistic regime in the strongly-driven TLS.} The gray box represents the regimes discussed in this paper. Relativistic effects become important when the normalized vector potential $a_0$ becomes smaller than 1. In the TLS system considered here, a Keldysh parameter smaller than 0.007 is required to enter the relativistic regime. } 
		\end{center}
	\end{figure}
	The classical equation of motion of a free electron in an optical field is given by
	\begin{align}
	\frac{\text{d} \textbf{p}}{\text{d} t} = -e \left(\textbf{E}(\textbf{x},t) + \textbf{v} \times \textbf{B}(\textbf{x}, t)\right),
	\label{equ. motion}
	\end{align}
	with peak magnetic field $|B_0| = E_0/c$ (c: speed of light, e: electron charge, p: electron momentum, v: electron velocity). In classical Newtonian mechanics we assume that $v \ll c$. This implies that the contribution of the second term in Eq.\,\eqref{equ. motion} is negligible. Assuming an oscillating electric field, the maximal velocity is $v_\text{max} = eE_0/(\omega m_e)$. When $v_\text{max}$ becomes comparable to $c$, the normalized vector potential 
	\begin{align}
	a_0=\frac{eE_0}{\omega m_e c},
	\end{align}
	reaches unity and relativistic effects become important \cite{Fuchs2009}. In case of a TLS, we replace the electron rest mass $m_e$ with the effective mass [Eq.\,\eqref{Eq. mass}] and obtain 
	\begin{align}
	a_{0, \text{TLS}} = \frac{v_\text{F}}{\gamma c}. 
	\end{align}
	The relativistic regime ($a_0>1$) is entered for $\gamma < \frac{v_F}{c}$. Hence, in the TLS considered here with $v_F=1$\,nm/fs, the magnetic component of the optical field can be neglected for $\gamma>0.007$. Figure\,\ref{S5} shows the full $\gamma$-M map including the relativistic regime.

	\begin{acknowledgements}
		This work has been supported in part by the European Research Council (Consolidator Grant “NearFieldAtto”), Deutsche Forschungsgemeinschaft (Sonderforschungsbereich 953 “Synthetic Carbon Allotropes”, project 182849149) and the PETACom project financed by Future and Emerging Technologies Open H2020 program. P. H. greatefully acknowledges a Fellowship from Max Planck Institute of the Science of Light (MPL). We thank Vladislav S. Yakovlev for discussions. 
	\end{acknowledgements}
	
	%
	
	
\end{document}